\documentclass[a4paper,12pt]{article}

\usepackage{a4wide}

\providecommand{\preprintno}[1]{\relax}
\usepackage{fancyhdr}
\def\preprint#1{\gdef\thepreprint{#1}}
\def\thepreprint{}
\fancypagestyle{fancyplain}{
  \fancyhf{}
  \fancyhead[R]{\thepreprint \\\hfill\\ February 1, 2006}
  
  \addtolength{\headsep}{2cm}
}
\usepackage{amsmath}
\usepackage{cite}
\usepackage{mcite}
\usepackage{hyperref}
\makeatletter
\renewcommand\section{\@startsection {section}{1}{\z@}%
                                   {-3.5ex \@plus -1ex \@minus -.2ex}%
                                   {2.3ex \@plus.2ex}%
                                   {\large\bf}}   
\renewcommand\subsection{\@startsection {subsection}{1}{\z@}%
                                   {-3.5ex \@plus -1ex \@minus -.2ex}%
                                   {2.3ex \@plus.2ex}%
                                   {\normalfont\bf}}         
\makeatother
\allowdisplaybreaks
\def\bra#1{\mathinner{\langle{#1}|}}
\def\ket#1{\mathinner{|{#1}\rangle}}
\def\braket#1{\mathinner{\langle{#1}\rangle}}
\newcommand{\vev}[1]{\braket{0|#1|0}}
\makeatletter
\newcommand{\fmslash}[2][0mu]{%
  \mathchoice
    {\fmsl@sh\displaystyle{#1}{#2}}%
    {\fmsl@sh\textstyle{#1}{#2}}%
    {\fmsl@sh\scriptstyle{#1}{#2}}%
    {\fmsl@sh\scriptscriptstyle{#1}{#2}}}
\newcommand{\fmsl@sh}[3]{%
  \m@th\ooalign{$\hfil#1\mkern#2/\hfil$\crcr$#1#3$}}
\makeatother
\newcommand{\sbraket}[1]{\lbrack #1\rbrack}
\newcommand{\Qsusy}{Q_{\text{\tiny SUSY}}}
\newcommand{\bq}{\begin{eqnarray}}
\newcommand{\eq}{\end{eqnarray}}
\renewcommand{\l}{\langle}
\renewcommand{\r}{\rangle} 
\newcommand{\eps}{\varepsilon}

\numberwithin{equation}{section}
\numberwithin{table}{section}
\numberwithin{figure}{section}

\begin{document}
%%%%%%%%%%%%%%%%%%%%%%%%%%%%%%%%%%%%%%%%%%%%%%%%%%%%%%%%%%%%%%%%%%%%%%%%
\preprint{MZ-TH/06-01 \\ PITHA 06/01}
\title{ SUSY Ward identities for multi-gluon 
helicity amplitudes with massive quarks }
\author{ 
Christian Schwinn\thanks{schwinn@physik.rwth-aachen.de}\\
{\normalsize \it  Institut f\"ur Theoretische Physik E}\\ 
{\normalsize \it RWTH Aachen, D - 52056 Aachen, Germany}\\\hfill\\
Stefan Weinzierl\thanks{stefanw@thep.physik.uni-mainz.de}\\
{\normalsize\it Institut f\"ur Physik, Johannes-Gutenberg-Universit\"at}\\
{\normalsize\it Staudingerweg 7,  D-55099 Mainz, Germany}
}
\setcounter{tocdepth}{2}
\setcounter{secnumdepth}{3}
 \date{}
\maketitle
\thispagestyle{fancyplain}
\setcounter{page}{0}
\begin{abstract}
We use supersymmetric Ward identities to relate multi-gluon helicity amplitudes
involving a pair of massive quarks to amplitudes with massive scalars.
This allows to use the recent results for scalar amplitudes with an 
arbitrary number of gluons obtained by on-shell recursion relations to 
obtain scattering amplitudes involving top quarks.  

\end{abstract}
\newpage
\setlength{\headheight}{\baselineskip}
\section{Introduction}
Recently, a number of efficient methods for the calculation of
helicity amplitudes in QCD have been introduced, motivated by the
relationship of QCD amplitudes to twistor string
theory~\cite{Witten:2003nn}.  In the Cachazo~-~Svr\v{c}ek~-~Witten
(CSW) construction~\cite{Cachazo:2004kj}, tree level QCD amplitudes
are constructed from vertices that are off-shell continuations of
maximal helicity violating (MHV) amplitudes~\cite{Parke:1986gb},
connected by scalar propagators.  
This formalism has been extended to massless
 quarks~\cite{Georgiou:2004wu,*Georgiou:2004by,*Wu:2004fb,*Wu:2004jx}
and loop diagrams~\cite{Brandhuber:2004yw,*Bena:2004xu}.
Subsequently a set of recursion
relations has been found~\cite{Britto:2004ap,*Britto:2005fq} that
involve only on-shell amplitudes with shifted, complex external
momenta.  
These methods extend 
earlier recursive formulations~\cite{Berends:1987me,Kosower:1989xy,Caravaglios:1995cd,*Kanaki:2000ey,*Moretti:2001zz}
which are based on currents with one off-shell leg.  
Recently a derivation of the CSW rules from the on-shell recursion
relations has been given~\cite{Risager:2005vk} and proofs of
the  on-shell
recursion relations using conventional Feynman
diagrams have appeared~\cite{Draggiotis:2005wq,Vaman:2005dt}.
Another elegant technique applicable to scattering amplitudes with massless particles
is the use of supersymmetric Ward
identities~(SWIs)~\cite{Grisaru:1976vm,*Grisaru:1977px,Parke:1985pn,Reuter:2002gn} to relate
helicity amplitudes involving massless quarks to those involving
massless scalars or to purely gluonic
ones~\cite{Parke:1985pn,Kunszt:1985mg,Morgan:1995te,*Chalmers:1997ui,Bidder:2005in}.

In the light of the major role played by top quark physics at the LHC, 
it would be desirable to extend such methods also to multi-parton
processes involving massive quarks. For instance, top-pair production
with one or two additional jets contributes to the
 background for Higgs production
by vector boson fusion or in association with a top
quark pair~\cite{Rainwater:2002hm}.
Tree amplitudes with a top pair and additional jets
are also required for the  next-to leading order
 corrections to top pair plus jet production~\cite{Brandenburg:2004fw}. 
Helicity amplitudes for a top pair plus three gluons can be found in~\cite{Bernreuther:2004jv}.
Tree level helicity amplitudes with massive quarks are also needed within the unitarity method~\cite{Bern:1994cg}
for one-loop amplitudes, where a heavy quark is circulating in the loop.
In this context, four point amplitudes have been
computed in~\cite{Bern:1995db} and four and five point amplitudes have
been obtained from on-shell recursion relations
in~\cite{Quigley:2005cu}.

The on-shell recursion relations are  also applicable to massive 
particles~\cite{Badger:2005zh,Badger:2005jv} and closed expressions
for some sets of amplitudes with massive scalars and
an arbitrary number of gluons have recently been 
found~\cite{Forde:2005ue}. In the MHV approach it has been possible
to include single external
massive gauge bosons or 
Higgs bosons~\cite{Dixon:2004za,*Badger:2004ty,*Bern:2004ba}.
For massive quarks, the MHV approach cannot be used in a straightforward
way since there are nonvanishing helicity amplitudes that go to zero
in the massless limit  where less than two
partons have a helicity opposite to the remaining ones.
In the on-shell approach only four point functions with massive quarks
have been computed so far~\cite{Badger:2005jv}.

As a first step to extend the formalism of~\cite{Cachazo:2004kj}
towards massive quarks,  
we have reformulated in~\cite{Schwinn:2005pi,Schwinn:2005zm} the Feynman rules of QCD including
massive quarks so that only scalar propagators  and a set of
cubic and quartic primitive vertices without Lorentz or spinor
indices appear. The cubic primitive vertices in this approach can also
 be used directly in the on-shell recursion relations. 

Recently, a particular set of multi-gluon helicity amplitudes with a pair of massive quarks
has been calculated in~\cite{Rodrigo:2005eu}.
These results should be related through supersymmetric relations to the one obtained in~\cite{Forde:2005ue}
for multi-gluon helicity amplitudes with a pair of massive scalars.
In this paper we work out the explicit transformation laws.
The derivation follows the lines of the original paper~\cite{Grisaru:1976vm,*Grisaru:1977px}.
There are several motivations for reconsidering the supersymmetric transformation laws:
First of all, by taking into account recent developments in the definition of off-shell polarization
spinors, we obtain short and elegant results.
Secondly, the results allow us to relate amplitudes with massive scalars to amplitudes with massive
quarks.
Since there are more results available for all-multiplicity amplitudes with massive scalars, this is useful
for top quark physics.
Finally, we are very careful to keep internally all signs correct. 
This is an issue since there are
different sign conventions in the supersymmetric and QCD community.
The final result can of course easily be related to any other sign convention.
Furthermore we have to respect the relative sign between amplitudes with anti-fermions and without.
We verify the correctness of the transformation laws by a direct computation of
amplitudes with a massive pair of quarks or scalars by using Berends-Giele type recursion relations and the
diagrammatic rules of~\cite{Schwinn:2005pi}.

This article is organized as follows. 
In section~\ref{sec:notation} we review the notation for colour ordered
partial amplitudes and the conventions for spinors and polarization vectors.
In section~\ref{sec:susy} we introduce our supersymmetric model, give the supersymmetric transformation
laws and apply them to derive relations among various amplitudes.
In section~\ref{sect:appl} we consider as an example amplitudes with a massive pair of quarks or scalars
plus an arbitrary number of positive helicity gluons.
Our conclusions are contained in section~\ref{sect:conlcusions}.

In appendix~\ref{sect:notation} we list our sign conventions.
Appendix~\ref{app:susy} contains the derivation of the supersymmetric transformation laws.
In appendix~\ref{sec:rules} we extend
the diagrammatic rules obtained in~\cite{Schwinn:2005pi} to the
supersymmetric model. 
In appendix~\ref{app:recursion} we give some technical details on the solution of the recurrence relation
discussed in section~\ref{sect:appl}.

% ----------------------------------------------------------------

\section{Review of notation and off-shell continuation}
\label{sec:notation}

\subsection{Colour decomposition}
In this work we are concerned with the evaluation of helicity
amplitudes with one massive quark pair. We employ the usual decomposition
of the full amplitude ${\cal A}_n$ into gauge invariant partial 
amplitudes $A_n$ defined 
by~\cite{Berends:1987me,Mangano:1990by,Dixon:1996wi}: 
\bq
\label{eq:color}
\lefteqn{
 {\cal A}_{n}(\bar{Q}_1,g_2,g_3,...,g_{n-1},Q_{n}) 
=  } & & \nonumber \\
 & & 
 g^{n-2} \sum_{\sigma\in S_{n-2}(2,...,n-1)}  \; \left(
 T^{a_{\sigma(2)}} ... T^{a_{\sigma(n-1)}} \right)_{ij}
 A_{n}\left(\bar{Q}_1, g_{\sigma(2)}, ..., g_{\sigma(n-1)}, Q_n \right), 
\eq
where the sum is over all  permutations of the external gluon legs.
The colour structure is contained in group theoretical factors given by
 traces over the generators
$T^a$ of the fundamental representation of $SU(3)$.

We will also encounter amplitudes with scalars transforming under
the fundamental representation of $SU(3)$. 
The colour decomposition is similar to the one in eq. (\ref{eq:color}):
\bq
\lefteqn{
 {\cal A}_{n}(\phi_1,g_2,g_3,...,g_{n-1},\phi_{n}) 
=  } & & \nonumber \\
 & & 
 g^{n-2} \sum_{\sigma\in S_{n-2}(2,...,n-1)}  \; \left(
 T^{a_{\sigma(2)}} ... T^{a_{\sigma(n-1)}} \right)_{ij}
 A_{n}\left(\phi_1, g_{\sigma(2)}, ..., g_{\sigma(n-1)}, \phi_n \right). 
\eq

\subsection{Polarization vectors and spinors}

It is a well known fact that any four-vector can be decomposed into a sum of two light-like four-vectors.
Given a fixed light-like four-vector $q$, one can associate to any four-vector $k$ 
(which need not be light-like) a light-like four-vector $k^\flat$, defined by
\begin{equation}
\label{eq:momentum}
  k^\mu = k^{\flat\mu}+\frac{k^2}{2(k\cdot q)} q^\mu.
\end{equation}
This prescription has also been used for the off-shell
continuation of MHV amplitudes~\cite{Bena:2004ry,*Kosower:2004yz} 
in the context of the CSW rules.
Unless stated otherwise, we will use the
same reference vector $q$ for all momenta.
The projected four-vector $k^\flat$ can be used to define continuations off the null-space of spinor products
as follows:
\begin{equation}
  \braket{k_1 k_2}=\braket{k_1^\flat-|k_2^\flat+},\qquad \qquad 
  \sbraket{k_1 k_2 }=\braket{k_1^\flat+|k_2^\flat-}.
\end{equation}
The two two-component spinors $\ket{k^\flat+}$ and
$\ket{k^\flat -}$ are the Weyl spinors corresponding to 
the projected momentum
$k^{\flat\mu}$. 
These can be used to define the off-shell continuation of the polarization vectors for
gluons:
\begin{equation}
\label{eq:os-pol}
  \epsilon^\pm_\mu(k,q)=\pm\frac{\braket{q\mp|\gamma_\mu|k^\flat\mp}}
  {\sqrt 2 \braket{q\mp| k^\flat\pm}}.
\end{equation}
Helicity methods can be extended to massive fermions~\cite{Kleiss:1985yh,Ballestrero:1994jn,*Dittmaier:1998nn,*vanderHeide:2000fx,Morgan:1995te,*Chalmers:1997ui,*Chalmers:1998jb,Rodrigo:2005eu}. 
We introduce an off-shell
continuation of massive spinors, using the projection~\eqref{eq:momentum} as for the
gluon polarization vectors:
\begin{equation}\label{eq:os-spinors}
  \begin{aligned}
  u(k,\pm)&=\frac{\fmslash k+m}{\braket{k^\flat\mp| q\pm}}\ket{q\pm}
 =\ket{k^\flat\mp}+\frac{m}{\braket{k^\flat\mp| q\pm}}\ket{q\pm},\\
 \bar u(k,\pm)&=\bra{q\mp}\frac{\fmslash k+m}{\braket{q\mp |k^\flat\pm}}
= \bra{k^\flat \pm}+\frac{m}{\braket{q\mp| k^\flat\pm}}\bra{q\mp},\\
  v(k,\pm)&=\frac{\fmslash k-m}{\braket{k^\flat\mp| q\pm}}\ket{q\pm}
 =\ket{k^\flat\mp}-\frac{m}{\braket{k^\flat\mp| q\pm}}\ket{q\pm},\\
 \bar v(k,\pm)&=\bra{q\mp}\frac{\fmslash k-m}{\braket{q\mp |k^\flat\pm}}
= \bra{k^\flat \pm}-\frac{m}{\braket{q\mp| k^\flat\pm}}\bra{q\mp}.
 \end{aligned}
\end{equation}
The normalization is chosen in order to allow for a smooth massless
limit. 
We label the helicities as if all particles were outgoing. 
As a consequence, the spinors $u(k)$ and $\bar{v}(k)$, which correspond
to particles with incoming momentum, have a reversed helicity assignment.

The reference momentum used in the definition~\eqref{eq:os-spinors} 
is not an unphysical quantity
that has to drop out in the final result for the helicity amplitudes, as
in the case of the definition of the gluon polarization vectors, 
but rather defines the quantization axis of the quark
spin~\cite{Kleiss:1985yh}.
Since for any given choice of $q$ the spinors~\eqref{eq:os-spinors} form 
a complete basis of solutions of the Dirac equation, the calculation can be simplified 
without loss of generality  by a suitable choice of $q$. 
The amplitudes for a desired physical polarization can be obtained by a straightforward linear transformation.

%%%%%%%%%%%%%%%%%%%%%%%%%%%%%%%%%%%%%%%%%%%%%%%%%%%%%%%%%%%%%%%%%%%%%%%%%%%%
\section{SUSY Ward identities for massive quarks}
\label{sec:susy}
Calculating examples of 
helicity amplitudes for a massive quark pair and positive
helicity gluons, one observes a 
simple relation to amplitudes with massive scalars that have been calculated
in~\cite{Bern:1996ja,Badger:2005zh,Forde:2005ue}. As a simple example, 
we find the four point function
\begin{equation}\label{eq:ttgg}
  A(\bar{Q}_1^+,g_2^+,g_3^+,Q_4^-)= 
\frac{\braket{4q}\sbraket{23}}{\braket{1q}\braket{23} } \frac{2 i m^2}{(k_1+k_2)^2-m^2}.
\end{equation}
The results for an arbitrary number of positive helicity gluons are quoted
in appendix~\ref{app:recursion}.
Comparing to the scattering amplitude
 for two massive scalars and two positive helicity 
gluons~\cite{Bern:1996ja,Badger:2005zh,Forde:2005ue} 
\begin{equation}\label{eq:ssgg}
  A(\bar{\phi}_1^+,g_2^+,g_3^+,\phi_4^-)=  
\frac{\sbraket{23}}{\braket{23} } \frac{2 i m^2}{(k_1+k_2)^2-m^2},
\end{equation}
we see these amplitudes are related by a 
factor $\frac{\braket{1q}}{\braket{4q}}$, similar to the SUSY
relation between quark and gluon amplitudes.
In the next subsection we define a supersymmetric toy model, which contains QCD as a subset, and derive
the supersymmetric transformation laws.
In subsection~\ref{sec:swi} we apply these relations to amplitudes involving a pair of massive quarks
or scalars.
It turns out, that the relations are particular simple, if the reference spinor defining the spin axis of the heavy
quarks, is chosen to be same for both the quark and the anti-quark.
In subsection~\ref{app:rodrigo} we discuss modifications which arise if the reference spinors are not identical.

\subsection{Definition of the model}
\label{sec:model}

To embed QCD with massive quarks into a supersymmetric extension, we have to use two
chiral super-multiplets $\Phi_+=(\varphi_+,\psi_+,F_+)$ and $\bar{\Phi}_-=(\bar{\varphi}_-,\bar{\psi}_-,\bar{F}_-)$
and combine the spinor component fields into Dirac spinors as follows 
\bq
 \bar{\Psi} = \left( \psi_-, \bar{\psi}_+ \right),
 & &
 \Psi = \left( \begin{array}{c} \psi_+ \\ \bar{\psi}_- \\ \end{array} \right).
\eq
The chiral fields are written in component fields as follows:
\bq
 \Phi_+(y,\theta) = \varphi_+(y) + \sqrt{2} \theta \psi_+(y) + \theta^2 F_+(y),
 & &
  y^\mu = x^\mu - i \theta \sigma^\mu \bar{\theta},
 \nonumber \\
 \bar{\Phi}_-(\bar{y},\bar{\theta}) = \bar{\varphi}_-(\bar{y}) + \sqrt{2} \bar{\theta} \bar{\psi}_-(\bar{y}) + \bar{\theta}^2 \bar{F}_-(\bar{y}),
 & &
 \bar{y}^\mu = x^\mu - i \bar{\theta} \bar{\sigma}^\mu \theta.
\eq
The Lagrange density of our model is given by
\bq
\label{eq:Lagrangian_1}
 {\cal L} & = & 
 \frac{1}{8 g^2} \; \mbox{Tr} \; \left. W W \right|_F
 + \frac{1}{8 g^2} \; \mbox{Tr} \; \left. \bar{W} \bar{W} \right|_{\bar{F}}
 \nonumber \\
 & &
 + \left. \bar{\Phi}_+ e^{-2 g V} \Phi_+ \right|_D
 + \left. \bar{\Phi}_- e^{2 g V^T} \Phi_- \right|_D
 + \left. m \Phi_- \Phi_+ \right|_F
 + \left. m \bar{\Phi}_+ \bar{\Phi}_- \right|_{\bar{F}},
\eq
where the vector multiplet $V^a=(A^a,\lambda^a,D^a)$ contains the gluon $A^a$ and the gluino $\lambda^a$.
The field strength is defined by
\bq
 W_A = - \frac{1}{4} \left( \bar{D} \bar{D} \right) \left( e^{2 g V} D_A e^{-2 g V} \right),
 \;\;\;
 \bar{W}_{\dot{A}} = - \frac{1}{4} \left( D D \right) \left( \left( \bar{D}_{\dot{A}} e^{-2 g V} \right) e^{2 g V} \right),
 \;\;\;
 V = T^a V^a,
\eq
and $V^a$ is given in the Wess-Zumino gauge by
\bq
 V^a & = & 
  \left( \theta \sigma^\mu \bar{\theta} \right) A^a_\mu 
 + i \theta^2 \left( \bar{\theta} \bar{\lambda}^a \right)
 - i \bar{\theta}^2 \left( \theta \lambda^a \right)
 + \frac{1}{2} \theta^2 \bar{\theta}^2 D^a.
\eq
For the gluino field we define a four-component spinor as follows:
\bq
 \bar{\Lambda} = \left( i \lambda^A, -i \bar{\lambda}_{\dot{A}} \right),
 & &
 \Lambda = \left( \begin{array}{c} i \lambda_A \\ -i \bar{\lambda}^{\dot{A}} \\ \end{array} \right).
\eq
It is also convenient to redefine the scalar fields as follows:
\bq
\bar{\phi}_- = \varphi_-,
 \;\;\;
\phi_+ = \bar{\varphi}_-,
 \;\;\;
\bar{\phi}_+ = \bar{\varphi}_+,
 \;\;\;
\phi_- = \varphi_+.
\eq
Eliminating the $D$- and $F$-terms and with the convention that
\bq
 \left( \bar{\phi}_\pm \right)^\dagger & = & \phi_\mp
\eq
we obtain
\bq
\label{eq:Lagrangian_2}
 {\cal L} & = & 
 - \frac{1}{4} \left( F^a_{\mu\nu} \right)^2 + \frac{i}{2} \bar{\Lambda}^a \gamma^\mu D^{ab}_\mu \Lambda^b 
 + i \bar{\Psi} \gamma^\mu D_\mu \Psi
 - m \bar{\Psi} \Psi
 \\
 & &
 +
 \left( D_\mu \phi_+ \right)^\dagger \left( D^\mu \phi_+ \right) 
 - m^2 \bar{\phi}_- \phi_+ 
 +
 \left( D_\mu \phi_- \right)^\dagger \left( D^\mu \phi_- \right) 
 - m^2 \bar{\phi}_+ \phi_- 
 \nonumber \\
 & &
 - \sqrt{2} g \left[ \bar{\phi}_+ \bar{\Lambda}^a T^a P_+ \Psi + \bar{\Psi} P_- \Lambda^a T^a \phi_- 
                    -\bar{\phi}_- \bar{\Lambda}^a T^a P_- \Psi - \bar{\Psi} P_+ \Lambda^a T^a \phi_+ \right]
 \nonumber \\
 & &
 - \frac{1}{2} g^2 \left( \bar{\phi}_+ T^a \phi_- -  \bar{\phi}_- T^a \phi_+ 
                     \right)^2,
 \nonumber 
\eq
where the covariant derivative in the fundamental and adjoint representation is given by
\bq
 D_\mu = \partial_\mu - i g T^a A^a_\mu,
 & &
 D^{ab}_\mu = \partial_\mu - g f^{abc} A^c_\mu.
\eq
The chiral projectors are denoted as usual by $P_\pm = 1/2(1\pm \gamma_5)$.
As in~\cite{Schwinn:2005pi} we can derive primitive vertices for the additional
interactions that arise compared to nonsupersymmetric QCD. The results are
collected in appendix~\ref{sec:rules}.

\subsection{Transformation of the fields}
\label{sec:massless-susy}

The SUSY transformations of helicity states have  been first derived
 in~\cite{Grisaru:1976vm,*Grisaru:1977px} and later applied to QCD helicity amplitudes~\cite{Parke:1985pn,Kunszt:1985mg}. 
We use a commuting SUSY generator 
$\Qsusy(\eta)=\eta^AQ_A + \bar{\eta}_{\dot A}\bar Q^{\dot A}$  
obtained by multiplying the SUSY generators
by a Grassmann valued spinor $\eta$. This generator satisfies the SUSY
algebra
\begin{equation}
  \label{eq:susy-algebra}
   \lbrack \Qsusy(\eta), \Qsusy(\xi)\rbrack=-2 i \bar \eta \fmslash \partial \xi.
\end{equation}
SUSY transformations of generic bosonic or fermionic fields 
$\Psi$ are generated by 
$\delta_\eta\Psi=\lbrack\Qsusy(\eta),\Psi\rbrack$.
With the notation
\begin{equation}
\label{eq:def-gamma}
  \Gamma^\pm_\eta(k) =\sqrt 2\braket{\eta\pm| k\mp},
 \;\;\;
  \Sigma_\eta^\pm(k,q)=\sqrt 2 
  m\frac{\braket{q\pm|\eta\mp}}{\braket{q\pm|k\mp}}
\end{equation}
one finds the transformation laws for outgoing particles: 
\bq
\label{eq:massive-susy}
 \delta_\eta \bar{Q}^\pm = \Gamma^\pm_\eta(k) \bar{\phi}^\pm - \Sigma^\mp_\eta(k,q) \bar{\phi}^\mp,
 & &
 \delta_\eta \bar{\phi}^\pm = \Gamma^\mp_\eta(k) \bar{Q}^\pm + \Sigma^\mp_\eta(k,q) \bar{Q}^\mp,
 \nonumber \\
  \delta_\eta Q^\pm =  -\Gamma^\pm_\eta(k) \phi^\pm + \Sigma^\mp_\eta(k,q) \phi^\mp,
 & &
 \delta_\eta \phi^\pm = -\Gamma^\mp_\eta(k) Q^\pm - \Sigma^\mp_\eta(k,q) Q^\mp.
\eq
Details on the derivation are given in appendix~\ref{app:susy}.
Eq.~\eqref{eq:massive-susy} is the main result of this paper and generalizes the well-known
results for massless particles to the massive case.
Compared to the massless case there are additional terms with $\Sigma^\pm$, which are proportional to the mass
of the particles in the super-multiplet.
For practical applications, one can achieve a considerable
simplification by noting that $\Sigma_q(k,q)=0$. Therefore the
additional contributions to~\eqref{eq:massive-susy} can be avoided
\emph{provided} the SUSY spinors $\ket{\eta\pm}$ are taken to be equal
to a Grassmann parameter $\theta$ times the reference spinors $\ket{q\pm}$ used in the definition of the massive
spinors~\eqref{eq:os-spinors}. 
This requires that the same reference spinors $\ket{q\pm}$ are chosen for all massive quarks in the amplitude.
In this case we have the following transformation laws, which are identical with the massless case:
\bq
\label{eq:massive-susy-short}
 \delta_\eta \bar{Q}^\pm = \Gamma^\pm_\eta(k) \bar{\phi}^\pm,
 & &
 \delta_\eta \bar{\phi}^\pm = \Gamma^\mp_\eta(k) \bar{Q}^\pm,
 \nonumber \\
  \delta_\eta Q^\pm =  -\Gamma^\pm_\eta(k) \phi^\pm,
 & &
 \delta_\eta \phi^\pm = -\Gamma^\mp_\eta(k) Q^\pm.
\eq
The gluon and the gluino transform as follows:
\bq
\label{eq:susy-gluon}
 & &
 \delta_\eta g^\pm 
  = - \Gamma^\pm_\eta(k) \bar{\Lambda}^\pm 
  = \Gamma^\pm_\eta(k) \Lambda^\pm,
 \nonumber \\
 & &
 \delta_\eta \bar{\Lambda}^\pm = -\Gamma^\mp_\eta(k) g^\pm,
 \;\;\;
 \delta_\eta \Lambda^\pm = \Gamma^\mp_\eta(k) g^\pm.
\eq

\subsection{SUSY Ward identities for massive quarks and scalars}
\label{sec:swi}

We can now derive relations connecting
amplitudes of massive quarks to amplitudes of massive scalars, generalizing
the example of the four point amplitudes~\eqref{eq:ttgg} and~\eqref{eq:ssgg}.  
In this section we will always choose $\eta=\theta q$ and use the same reference vector $q$
in the definition of the spinors for all massive quarks. Therefore terms proportional to
$\Sigma^\pm$ do not contribute. 
Furthermore we treat in this section all particles as outgoing.

Since the SUSY charge $Q_{\text{SUSY}}$ annihilates the vacuum for unbroken
SUSY, we obtain SUSY Ward Identities (SWIs) from the relation~\cite{Grisaru:1976vm}
\begin{equation}
\label{eq:swi}
  0=\vev{\lbrack Q_{\text{SUSY}}(\eta), z_1\dots z_n\rbrack}
=\sum _i\vev{z_1 \dots \lbrack Q_{\text{SUSY}}(\eta), z_i\rbrack \dots z_n},
\end{equation}
where the $z_i$ are arbitrary creation or annihilation operators.

Applying the SUSY transformation to the amplitude 
\bq
 A(\bar{\phi}_1^+, g_2^+, ..., g_{n-1}^+, Q_n^- )
\eq
with one $\bar{\phi}^+$ scalar, $(n-2)$ positive helicity
gluons and a negative helicity quark results in
the relation
 \begin{multline}
\label{eq:swi-derive}
\Gamma^-_{\eta}(k_1)A({\bar Q}_1^+,g_2^+,\dots, g_{n-1}^+, Q_n^-)
-\sum_{i=2}^{n-1}
\Gamma_{\eta}^+(k_i) A(\bar{\phi}_1^+, g_2^+,\dots, \bar{\Lambda}_i^+,\dots, g_{n-1}^+, Q_n^-)\\
-\Gamma_{\eta}^-(k_n)A(\bar{\phi}_1^+,g_2^+,\dots, g_{n-1}^+, \phi_n^-) = 0.
 \end{multline}
Note that these identities hold to all orders in perturbation theory in
the supersymmetric toy model. However, only on tree level the scalar
and quark amplitudes are identical to those in pure QCD while in
higher orders loops involving SUSY particles contribute.  

The terms
arising from the transformations of the gluons vanish at tree level
since there are no primitive vertices of the form $V(\bar{\phi}_1^+,
\bar{\Lambda}_i^+,Q_n^-)$.  Therefore we obtain the relation
\begin{equation}
 \label{eq:massive-swi}
   A_n({\bar Q}_1^+,g_2^+,\dots, g_{n-1}^+, Q_n^-)=
\frac{\braket{nq}}{\braket{1q}} A_n(\bar{\phi}_1^+,g_2^+,\dots, g_{n-1}^+, \phi_n^-),
\end{equation}
which generalizes the relation among the four point amplitudes~\eqref{eq:ttgg}
and~\eqref{eq:ssgg} to an arbitrary number of positive helicity gluons.
In the same way, the SUSY transformation of 
$A(\bar{Q}_1^-, g_2^+, ..., g_{n-1}^+, \phi_n^+ )$ gives the corresponding relation
for exchanged quark helicities:
\begin{equation}
\label{eq:massive-swi-bar}
  A_n(\bar{Q}_1^-, g_2^+, ..., g_{n-1}^+, Q_n^+)=
-\frac{\braket{1q}}{\braket{nq}} A_n(\bar{\phi}_1^-, g_2^+, ..., g_{n-1}^+, \phi_n^+).
\end{equation}

We can also derive SWIs for amplitudes including a negative
helicity gluon. Leaving the helicity of one massive quark arbitrary, these
relations read
\begin{subequations}
\label{eq:neg-swi}
 \begin{multline}
0= \delta_{\eta} A(\bar{\phi}_1^+,g_2^+,\dots,g_j^-,\dots, Q_n^\pm)=
\Gamma^-_{\eta}(k_1)A(\bar{Q}_1^+,g_2^+,\dots,g_j^-,\dots, Q_n^\pm)\\
-\Gamma^-_{\eta}(k_j)A(\bar{\phi}_1^+,g_2^+,\dots,\bar{\Lambda}_j^-,\dots, Q_n^\pm)
-\Gamma_{\eta}^\pm(k_n)
A(\bar{\phi}_1^+,g_2^+,\dots,g_j^-,\dots, \phi_n^\pm),
 \end{multline}
where the terms from the transformation of the positive helicity
gluons vanish for the same reason as above.  For the amplitude with
two positive helicity quarks $A(\bar{Q}_1^+,\dots,g_j^-,\dots, Q_n^+)$,
the last term vanishes since it involves two different scalars $\bar{\phi}^+$ and $\phi^+$. 
This `helicity flip' amplitude is therefore directly related
to an amplitude involving a gluino and a scalar.
\bq
A(\bar{Q}_1^+,g_2^+,\dots,g_j^-,\dots, Q_n^+)
 & = & 
 \frac{\l j q \r}{\l 1 q \r} A(\bar{\phi}_1^+,g_2^+,\dots,\bar{\Lambda}_j^-,\dots, Q_n^+).
\eq
For the amplitude with one
negative helicity quark, all three contributions survive.  
The corresponding identity for the opposite quark helicities is  given by
\begin{multline}
0= \delta_{\eta} A(\bar{Q}_1^\pm,g_2^+,\dots,g_j^-,\dots, \phi_n^+)=
\Gamma^\pm_{\eta}(k_1)A(\bar{\phi}_1^\pm,g_2^+,\dots,g_j^-,\dots, \phi_n^+)\\
-\Gamma^-_{\eta}(k_j)A(\bar{Q}_1^\pm,g_2^+,\dots,\Lambda_j^-,\dots, \phi_n^+)
+ \Gamma_{\eta}^-(k_n)A(\bar{Q}_1^\pm,g_2^+,\dots,g_j^-,\dots, Q_n^+).
\end{multline}
\end{subequations}

We can simplify the identities~\eqref{eq:neg-swi} further
by a specific choice of the reference spinors.
With the choice $\ket{q+}=\ket{j+}$ we obtain 
\begin{align}
\label{eq:neg-swi1}
\left.A(\bar{Q}_1^+,g_2^+,\dots,g_j^-,\dots, Q_n^+)\right|_{\ket{q+}=\ket{j+}}&=0,\\
\left.A(\bar{Q}_1^+,g_2^+,\dots,g_j^-,\dots, Q_n^-)\right|_{\ket{q+}=\ket{j+}}&=
\frac{\braket{nj}}{\braket{1j}}  
A_n(\bar{\phi}_1^+,g_2^+,\dots,g_j^-,\dots, \phi_n^-),
\label{eq:neg-swi2}\\
\left.A(\bar{Q}_1^-,g_2^+,\dots,g_j^-,\dots, Q_n^+)\right|_{\ket{q+}=\ket{j+}}&=
-\frac{\braket{1j}}{\braket{nj}}  
A_n(\bar{\phi}_1^-,g_2^+,\dots,g_j^-,\dots, \phi_n^+).
\end{align}
The scalar amplitudes appearing in these identities have been computed
for an arbitrary number of gluons in~\cite{Forde:2005ue} using on-shell
recursion relations. 
However it should be noted that the choice $\ket{q+}=\ket{j+}$ defines the quantization axis 
of the spins of the heavy quarks.

Finally we can consider amplitudes with two negative helicity quarks
and positive helicity gluons only. In this case the SWI relates
the quark amplitude to a sum of gluino amplitudes:
\begin{multline}
  0= \delta_{\eta} A(\bar{\phi}_1^-, g_2^+, ..., g_{n-1}^+, Q_n^-)=
  \Gamma^+_{\eta}(k_1)A(\bar{Q}_1^-, g_2^+, ..., g_{n-1}^+, Q_n^-)\\
  -\sum_{i=2}^{n-1} \Gamma^+_{\eta}(k_i)
  A(\bar{\phi}_1^-, g_2^+,\dots, \bar{\Lambda}_i^+,\dots, g_{n-1}^+, Q_n^-),
\end{multline}
where the additional term from the transformation of the quark vanishes.

These last examples show
that SUSY methods can be extended to amplitudes with additional negative helicity
partons. However in general these relations will involve amplitudes with external gluinos.

%%%%%%%%%%%%%%%%%%%%%%%%%%%%%%%%%%%%%%%%%%%%%%%%%%%%%%%%%%%%%%%%%%%%%%

\subsection{Alternative choice of reference spinors} 
\label{app:rodrigo}
In~\cite{Rodrigo:2005eu} a different choice of auxiliary spinors
than in~\eqref{eq:os-spinors} has been used to calculate
helicity amplitudes for a massive quark pair and
an arbitrary number of
 positive helicity gluons using the Berends-Giele relations.
In this subsection we show how to relate our results to those
of~\cite{Rodrigo:2005eu}.
In that work the  momenta of the massive
quarks  $k_1$ and $k_n$ are decomposed in terms
of two lightlike vectors
$\widehat k_{1/n}$ according to
\begin{equation}
\label{eq:rodrigo-momenta}
 k_1^\mu =\beta_+ \widehat k_1^\mu +
  \beta_- \widehat k_n^\mu,\quad\quad
   k_n^\mu =\beta_- \widehat k_1^\mu +
  \beta_+ \widehat k_n^\mu,
\end{equation}
with $\beta_\pm=\tfrac{1}{2}(1\pm\beta)$ where $\beta=\sqrt{1-\frac{4m^2}{(k_1+k_n)^2}}$ is the quark velocity.
The spinors are then defined  similarly 
to~\eqref{eq:os-spinors} but using 
$\beta_+ \hat k_n$ as reference momentum for $\bar{Q}_1$ and $\beta_+ \hat k_1$ as reference
momentum for $Q_n$. Explicitly they read:
\begin{equation}
\label{eq:rodrigo-spinors}
  \begin{aligned}
   \bar u(1,\pm)
    &=\bra{\widehat n\mp}
    \frac{\fmslash k_1+m}{\beta_+^{1/2}\braket{\widehat n\mp |\widehat 1\pm}}
    =\beta_+^{1/2}\bra{\widehat 1 \pm}
    +\frac{m}{\beta_+^{1/2}\braket{\widehat n\mp |\widehat 1\pm}} 
    \bra{\widehat n\mp}, \\
     v(n,\pm)&=
     \frac{\fmslash k_n-m}{\beta_+^{1/2}
       \braket{\widehat n\mp| \widehat 1\pm}}\ket{\widehat 1\pm}
     =\beta_+^{1/2} \ket{\widehat n \mp} -
     \frac{m}{\beta_+^{1/2}\braket{\widehat n \mp| \widehat 1\pm}}
     \ket{\widehat 1\pm}.
   \end{aligned}
\end{equation}
Since the choice of different reference spinors for the two quarks
corresponds to different spin-vectors, 
different helicity combinations give a nonvanishing result compared to
our formalism where the same reference spinor is used for both quarks.
The general transformations~\eqref{eq:massive-susy} still relate amplitudes with massive quarks to amplitudes with
massive scalars. As an example we consider the SWI obtained from
\bq
 \delta_\eta  A\left(\bar \phi_1^+, g_2^+, ..., g_{n-1}^+, Q_n^+\right) & = & 0.
\eq
The explicit forms of the 
relevant transformation formulae from~\eqref{eq:massive-susy} read:
\bq
 \delta_\eta Q_n^+ 
 & = &
- \sqrt{2} \beta_+^{1/2} \Bigl\l \eta + | \widehat n - \Bigr\r {\phi}_n^+ 
 + \sqrt{2} m \beta_+^{-1/2} \frac{\left\l \widehat 1 - | \eta + \right\r}{\left\l \widehat 1 - | \widehat n + \right\r} {\phi}_n^-,
 \nonumber \\
 \delta_\eta \bar\phi_1^+ & = & 
 \sqrt{2} \beta_+^{1/2} \Bigl\l \eta - | \widehat 1 + \Bigr \r \bar{Q}_1^+ 
 + \sqrt{2} m \beta_+^{-1/2} \frac{\Bigl\l \widehat n - | \eta + \Bigr\r}{\left\l \widehat n - | \widehat 1 + \right\r} \bar{Q}_1^-,
\eq
and by choosing $\eta=\theta \widehat k_n$ we obtain for the helicity flip amplitude 
\bq
\label{eq:swi-rodrigo}
 A\left(\bar Q_1^+, g_2^+, ..., g_{n-1}^+, Q_n^+\right) 
 & = & \frac{m}{\beta_+ \left\l \widehat 1 - | \widehat n + \right\r}
 A\left(\bar \phi_1^+, g_2^+, ..., g_{n-1}^+, \phi_n^-\right).
\eq
This relation can immediately be verified for $n=4$:
Ref~\cite{Rodrigo:2005eu} obtains for the four point function 
the result
\begin{equation}\label{eq:ttgg-rodrigo}
 A\left(\bar Q_1^+, g_2^+, g_3^+, Q_4^+\right) 
 = 
 \frac{m}{\beta_+ \left\l \widehat 1 - | \widehat 4 + \right\r}
\frac{\sbraket{23}}{\braket{23}} \frac{2 i m^2}{(k_1+k_2)^2-m^2}.
\end{equation}
The scalar amplitude is given by \cite{Forde:2005ue} 
\begin{equation}
 A\left(\bar \phi_1^+, g_2^+, g_3^+, \phi_4^-\right)
 = 
\frac{\sbraket{23}}{\braket{23} } \frac{2 i m^2}{(k_1+k_2)^2-m^2}.
\end{equation}

%%%%%%%%%%%%%%%%%%%%%%%%%%%%%%%%%%%%%%%%%%%%%%%%%%%%%%%%%%%%%%%%%%%%%%%%
\section{Recursion relation and results
 for amplitudes with positive helicity gluons}
\label{sect:appl}

In this section we consider the degree zero amplitudes
\bq
\label{allgluonampl}
A(\bar Q_1^+,g_2^+,\dots,g_{n-1}^+, Q_n^-) 
&
\mbox{and}
&
A(\bar \phi_1^+,g_2^+,\dots,g_{n-1}^+, \phi_n^-)
\eq 
with a massive pair of quarks or scalars
plus an arbitrary number of positive helicity gluons.
In the following we will discuss amplitudes with one particle off-shell, this particle will be marked by a hat.
It is also convenient to treat $Q_n$ as an incoming quark rather than an outgoing anti-quark.
Let us take the outgoing particles $1$ to $n-1$ with positive energy. Therefore $p_n=k_1+...+k_{n-1}=-k_n$ has
a positive energy component.
With our phase conventions, the corresponding amplitudes are related by
\bq
A(\bar Q_1^+,g_2^+,\dots,g_{n-1}^+, Q_{k_n}^-) 
 & = & i 
A(\bar Q_1^+,g_2^+,\dots,g_{n-1}^+| Q_{p_n=-k_n}^-).
\eq
In the following we will use the diagrammatic rules of~\cite{Schwinn:2005pi}
that are briefly reviewed in appendix~\ref{sec:rules}.
The amplitudes in eq.(\ref{allgluonampl}) are of degree zero,
therefore in our conventions no helicity-flip vertex
occurs. 
As a consequence, the corresponding amplitudes with one-leg off-shell satisfy rather
simple Berends-Giele type recurrence relations:
\begin{multline}
\label{eq:Qrecurs}
   A_n( \bar Q_1^+, ..., g_{n-1}^+|\widehat{Q}_{p_n}^-)
= \sum\limits_{j=2}^{n-1} 
 V_3(\bar Q_{1,j}^+,g_{j,n}^+, Q_{p_n}^-) \,
 \frac{i}{k_{1,j}^2-m^2} 
A_j(\bar Q_1^+,...,g_{j-1}^+|\widehat{Q}^-_{1,j})\, \\
 \frac{i}{k_{j,n}^2} \,A_{n-j+1}(g_j^+,...,g_{n-1}^+,\widehat{g}_{-(j,n)}^- ),
\end{multline}
where we have defined
\begin{equation}
   k_{i,j}  =  \sum\limits_{l=i}^{j-1} k_l.
\end{equation}
The corresponding recurrence relation for the scalars reads:
\begin{multline}
\label{eq:phirecurs}
   A_n( \bar \phi_1^+, ..., g_{n-1}^+, \widehat{\phi}_{n}^-)
= \sum\limits_{j=2}^{n-1} 
 V_3(\bar \phi_{1,j}^+,g_{j,n}^+, \phi_{n}^-) \,
 \frac{i}{k_{1,j}^2-m^2} 
A_j(\bar \phi_1^+,...,g_{j-1}^+, \widehat{\phi}^-_{-(1,j)})\, \\
 \frac{i}{k_{j,n}^2} \,A_{n-j+1}(g_j^+,...,g_{n-1}^+,\widehat{g}_{-(j,n)}^- ).
\end{multline}
Here the quartic primitive scalar-gluon vertices do not contribute since the off-shell continuation of the polarization vectors~\eqref{eq:os-pol} implies $\epsilon^+(k_i,q)\cdot \epsilon^+(k_j,q)=0$. 

The recursion starts with
the two-point amplitudes, which are given by the inverse
propagators:
\begin{equation}
  \begin{aligned}
   A_2(g_j^+, \widehat{g}_{-j}^- ) &=  - i k_{j}^2, & &
   A_2(\bar Q_1^+| \widehat{Q}_1 ) =  A_2(\bar \phi_1^+, \widehat{\phi}_{-1} ) &= - i (k_{1}^2-m^2).
  \end{aligned}
\end{equation}
These recurrence relations are solved explicitly in appendix~\ref{app:recursion}.
In the present example  
we can use the recursion relation to establish the 
relation~\eqref{eq:massive-swi} between
the amplitudes of quarks and scalars directly.

After applying crossing symmetry the quark amplitude entering the SWI
 can be written as
\begin{multline}
\label{eq:Qrecurs2}
   A_n( \bar Q_1^+, ..., g_{n-1}^+,\widehat{Q}_{n}^-)
=i    A_n( \bar Q_1^+, ..., g_{n-1}^+|\widehat{Q}_{p_n=-k_n}^-)\\
= i \sum\limits_{j=2}^{n-1} 
 V_3(\bar Q_{1,j}^+,g_{j,n}^+, Q_{p_n}^-) \,
 \frac{i}{k_{1,j}^2-m^2}\\ 
(-i) A_j(\bar Q_1^+,...,g_{j-1}^+,\widehat{Q}^-_{-(1,j)})\, 
 \frac{i}{k_{j,n}^2} \,A_{n-j+1}(g_j^+,...,g_{n-1}^+,\widehat{g}_{-(j,n)}^- ).
\end{multline}
We now use the induction assumption
 that the identity~\eqref{eq:massive-swi}
holds up to
$n-1$ already for the one-leg off-shell amplitudes.
This is true for $n=2$:
\bq
  A_2( \bar Q_1^+, \widehat{Q}_{2}^-) & = & i A_2( \bar \phi_1^+, \widehat{\phi}_{2}^-).
\eq
Since the primitive cubic scalar-gluon  vertex~\eqref{eq:susy_scalar}
and the quark gluon vertex~\eqref{eq:qqg} satisfy the identity
\begin{equation}
\label{eq:susy-vertex}
 V(\bar Q_k^+,l^+,Q_{p_n=-k_n}^-)= 
\frac{\braket{p_nq}}{\braket{kq}}  V(\bar \phi_{k}^+,l^+,\phi_{n}^-)
=-i \frac{\braket{nq}}{\braket{kq}}  V(\bar \phi_{k}^+,l^+,\phi_{n}^-),
 \end{equation}
we find that every term of the sum in the recursion relation
contains a product of the form
\begin{multline}
 V_3(\bar Q_{1,j}^+,g_{j,n}^+, Q_{-n}^-)A_j(\bar Q_1^+,\dots, g_{j-1}^+,
 \widehat {Q}_{-(1,j)}^-)\\
=\left(-i \frac{\braket{nq}}{\braket{(1,j)q}} V_3(\bar \phi_{1,j}^+,g_{j,n}^+,\phi_n^-)\right)
\left(
\frac{\braket{-(1,j)q}}{\braket{1q}}
A_j(\bar \phi_1^+,\dots (j-1)^+ ,\widehat \phi_{-(1,j)}^-)\right)\\
=\frac{\braket{nq}}{\braket{1q}} V_3(\bar \phi_{1,j}^+,g_{j,n}^+,\phi_n^-)
A_j(\bar \phi_1^+,\dots (j-1)^+ ,\widehat \phi_{-(1,j)}^-)
\end{multline}
Therefore in every term appears the same overall-factor 
$\frac{\braket{nq}}{\braket{1q}}$ that can be pulled out of the sum.
The remaining factors give the scalar amplitude~\eqref{eq:phirecurs}
and we obtain the SWI~\eqref{eq:massive-swi}:
\bq
A(\bar Q_1^+,g_2^+,\dots,g_{n-1}^+, Q_n^-) 
&=&
  \frac{\braket{n q}}{\braket{1 q}}
A(\bar \phi_1^+,g_2^+,\dots,g_{n-1}^+, \phi_n^-).
\eq 
This is a non-trivial cross-check on the correctness of the supersymmetric transformation laws in eq.(\ref{eq:massive-susy}), in particular since
the equivalence of the analytic expressions for the all-multiplicity
 quark amplitudes~\cite{Rodrigo:2005eu} 
with the scalar amplitudes~\cite{Forde:2005ue}
is not immediate.

%%%%%%%%%%%%%%%%%%%%%%%%%%%%%%%%%%%%%%%%%%%%%%%%%%%%%%%%%%%%%%%%%%%%%%%%%%%%

\section{Conclusions}
\label{sect:conlcusions}

In this paper we considered relations based on supersymmetry 
between multi-gluon amplitudes with a pair massive quarks and corresponding amplitudes
where the quarks are replaced by massive scalars.
These relations generalize the well-known formulae for the massless case.
The relations are useful as they allow to obtain amplitudes with heavy quarks
from amplitudes with scalars. 
Furthermore, if both sets of amplitudes are known, the supersymmetric relations provide
a valuable cross-check.

\section*{Note added in proof}

After the submission of this paper, a form of the scalar
and quark amplitudes with an arbitrary number of positive helicity
gluons has been found~\cite{Ferrario:2006np}  that is much simpler compared to
those given in~\cite{Forde:2005ue,Rodrigo:2005eu}
and in~\eqref{eq:a++}
 and that  has been
used to verify the
identity~\eqref{eq:swi-rodrigo} explicitly for an arbitrary number of
gluons.

\section*{Acknowledgments}
CS was supported by the DFG through the Graduiertenkolleg "Eichtheorien" at Mainz University and by the DFG Sonderforschungsbereich/Transregio 9 "Computergest\"utzte Theoretische Teilchenphysik".
%%%%%%%%%%%%%%%%%%%%%%%%%%%%%%%%%%%%%%%%%%%%%%%%%%%%%%%%%%%%%%%%%%%%%%%%%%%%
\appendix

\section{Notation and conventions}
\label{sect:notation}

In this appendix we give a comprehensive summary of our notation and sign conventions.
The convention for the metric tensor is
$g_{\mu \nu} = \mbox{diag}(+1,-1,-1,-1)$.
We use the
Weyl representation for the Dirac matrices
\bq
 \gamma^{\mu} = \left(\begin{array}{cc}
 0 & \sigma^{\mu} \\
 \bar{\sigma}^{\mu} & 0 \\
\end{array} \right),
& &
\gamma_{5} = i \gamma^0 \gamma^1 \gamma^2 \gamma^3 
           = \left(\begin{array}{cc}
 1 & 0 \\
 0 & -1 \\
\end{array} \right),
\eq
where
$\sigma_{A \dot{B}}^{\mu} = \left( 1 , - \vec{\sigma} \right)$,
$\bar{\sigma}^{\mu \dot{A} B} = \left( 1 ,  \vec{\sigma} \right)$
and $\vec{\sigma} = \left( \sigma_x, \sigma_y, \sigma_z \right)$ are the Pauli matrices.
The chiral projectors are as usual $P_\pm = 1/2(1 \pm \gamma_5 )$.
The two-dimensional antisymmetric tensor is defined by
\bq
\varepsilon^{AB} = \varepsilon^{\dot{A}\dot{B}} = \varepsilon_{AB} = \varepsilon_{\dot{A}\dot{B}}
 =
\left(\begin{array}{cc}
 0 & 1\\
 -1 & 0 \\
\end{array} \right).
\eq
Indices of two-component Weyl spinors are raised and lowered as follows:
\bq
 \psi^A = \varepsilon^{AB} \psi_B,
  \;\;\;
 \bar{\psi}^{\dot{A}} = \varepsilon^{\dot{A}\dot{B}} \bar{\psi}_{\dot{B}},
  \;\;\;
 \bar{\psi}_{\dot{B}} = \bar{\psi}^{\dot{A}} \varepsilon_{\dot{A}\dot{B}}, 
  \;\;\;
 \psi_B = \psi^A \varepsilon_{AB}.
\eq
The generators of supersymmetry are taken to be
\bq
 Q_A = \frac{\partial}{\partial \theta^A} + i \sigma^{\mu}_{A \dot{A}} \bar{\theta}^{\dot{A}} \partial_\mu, 
  & & 
 \bar{Q}^{\dot{A}} =  
 \frac{\partial}{\partial \bar{\theta}_{\dot{A}}} + i \bar{\sigma}^{\mu\;\dot{A} B} \theta_B \partial_\mu.
\eq
They satisfy the supersymmetry algebra:
\bq
 \left\{ Q_A, \bar{Q}_{\dot{A}} \right\} & = & - 2 i \sigma^\mu_{A \dot{A}} \partial_\mu.
\eq
Spinor products are defined as
\bq
\xi \eta = \xi^A \eta_A, & & \bar{\xi} \bar{\eta} = \bar{\xi}_{\dot{A}} \bar{\eta}^{\dot{A}}.
\eq
In the bra-ket notation spinor products are denoted as
\bq
 \braket{p q} = \braket{p - | q + } = p^A q_A,
 & &
 \sbraket{q p} = \braket{q + | p - } = q_{\dot{A}} p^{\dot{A}},
\eq 
In terms of the light-cone components 
\bq
p_+ = p_0 + p_3, \;\;\; p_- = p_0 - p_3, \;\;\; p_{\bot} = p_1 + i p_2,
                                         \;\;\; p_{\bot^\ast} = p_1 - i p_2.
\eq
of a null-vector $p^\mu$, the corresponding massless spinors $\l p \pm |$ and $| p \pm \r$ 
can be chosen as
\bq
\left| p+ \right\r = \frac{e^{-i \frac{\phi}{2}}}{\sqrt{\left| p_+ \right|}} \left( \begin{array}{c}
  -p_{\bot^\ast} \\ p_+ \end{array} \right),
 & &
\left| p- \right\r = \frac{e^{-i \frac{\phi}{2}}}{\sqrt{\left| p_+ \right|}} \left( \begin{array}{c}
  p_+ \\ p_\bot \end{array} \right),
 \nonumber \\
\left\l p+ \right| = \frac{e^{-i \frac{\phi}{2}}}{\sqrt{\left| p_+ \right|}} 
 \left( -p_\bot, p_+ \right),
 & &
\left\l p- \right| = \frac{e^{-i \frac{\phi}{2}}}{\sqrt{\left| p_+ \right|}} 
 \left( p_+, p_{\bot^\ast} \right),
\eq
where the phase $\phi$ is given by
\bq
p_+ & = & \left| p_+ \right| e^{i\phi}.
\eq
If $p_+$ is real and $p_+>0$ we have the following relations between a spinor corresponding to a vector $p$
and a spinor corresponding to a vector $(-p)$:
\bq
 \left| \left(-p\right) \pm \right\r & = & i \left| p \pm \right\r,
 \nonumber \\
 \left\l \left(-p\right) \pm \right| & = & i \left\l p \pm \right|.
\eq
Therefore the
spinors of massive quarks and anti-quarks are related by
$u(-k,\pm)=i v(k,\pm)$ and $\bar u(-k,\pm)=i \bar v(k,\pm)$. The
polarization vectors of the gluons are unchanged under the reversal of
the momentum.  Denoting a partial amplitude with $n$ incoming
particles with momenta $p_i$ and $m$ outgoing particles with momenta
$k_j$ by
\bq
  A_{n+m}(\Phi_{k_1}... \Phi_{k_m}|\Phi_{p_1}... \Phi_{p_n})
\eq
the amplitude with an outgoing anti-quark is related to that with an incoming 
quark by 
\bq
\label{eq:crossing}
 A(..,\, Q_{k_j},\,...|...)=-i    A(... |...,\,Q_{p_i=-k_j}, ... ) 
\eq
where the notation used for the quarks follows that of the spinors, i.e. 
incoming quarks and outgoing anti-quarks are denoted by $Q$ and outgoing quarks 
and incoming anti-quarks by $\bar Q$. The same phase as in~\eqref{eq:crossing} 
appears for the exchange of an outgoing quark with an incoming anti-quark while
for gluons and scalars no phase appears and it is sufficient to reverse the 
momentum.

%%%%%%%%%%%%%%%%%%%%%%%%%%%%%%%%%%%%%%%%%%%%%%%%%%%%%%%%%%%%%%%

\section{SUSY transformations of helicity states}
\label{app:susy}

In this appendix we give the derivation of the supersymmetric transformations in eq. (\ref{eq:massive-susy}) 
and eq. (\ref{eq:susy-gluon}).
We follow closely the approach of ref. \cite{Grisaru:1976vm}.
The Lagrangian in eq. (\ref{eq:Lagrangian_1}) is invariant under supersymmetric transformations.
The supersymmetric transformations act
on the component fields of $\Phi_+$ as follows:
\bq
 \delta_\eta \varphi_+ & = & \sqrt{2} \eta \psi_+, 
 \nonumber \\
 \delta_\eta \psi_+ & = & - \sqrt{2} i \sigma^\mu \bar{\eta} \partial_\mu \varphi_+ + \sqrt{2} \eta F_+,
 \nonumber \\
 \delta_\eta F_+ & = & - \sqrt{2} i \bar{\eta} \bar{\sigma}^\mu \partial_\mu \psi_+.
\eq
Here we used the two-component Weyl spinor notation.
The components of $\bar{\Phi}_-$ transform as
\bq
 \delta_\eta \bar{\varphi}_- & = & \sqrt{2} \bar{\eta} \bar{\psi}_-,
 \nonumber \\
 \delta_\eta \bar{\psi}_- & = & - \sqrt{2} i \bar{\sigma}^\mu \eta \partial_\mu \bar{\varphi}_-  
                   + \sqrt{2} \bar{\eta} \bar{F}_-,
 \nonumber \\
 \delta_\eta \bar{F}_- & = & - \sqrt{2} i \eta \sigma^\mu \partial_\mu \bar{\psi}_-.
\eq
Finally, the components of the vector multiplet transform as
\bq
 \delta_\eta {A^a_{\mu}} & = & 
 i \left( \bar{\eta} \bar{\sigma}^\mu \lambda^a - \bar{\lambda}^a \bar{\sigma}^\mu \eta \right),
 \nonumber \\
 \delta_\eta {\lambda^a} & = & \frac{1}{2} \left( \sigma^{\mu} \bar{\sigma}^\nu - \sigma^{\nu} \bar{\sigma}^\mu \right) \eta \partial_\mu A^a_\nu + i \eta D^a,
 \nonumber \\
 \delta_\eta {D^a} & = & \eta \sigma^\mu \partial_\mu \bar{\lambda}^a + \partial_\mu \lambda^a \sigma^\mu \bar{\eta}.
\eq
We recall the connection between the two-component Weyl spinor notation and the four-component Dirac spinor notation:
\bq
 \bar{\Psi} = \left( \psi_-, \bar{\psi}_+ \right),
 & &
 \Psi = \left( \begin{array}{c} \psi_+ \\ \bar{\psi}_- \\ \end{array} \right),
 \nonumber \\
 \bar{\Lambda} = \left( i \lambda^A, -i \bar{\lambda}_{\dot{A}} \right),
 & &
 \Lambda = \left( \begin{array}{c} i \lambda_A \\ -i \bar{\lambda}^{\dot{A}} \\ \end{array} \right).
\eq
For the scalars we used the redefinition 
\bq
\bar{\phi}_- = \varphi_-,
 \;\;\;
\phi_+ = \bar{\varphi}_-,
 \;\;\;
\bar{\phi}_+ = \bar{\varphi}_+,
 \;\;\;
\phi_- = \varphi_+.
\eq
together with the convention that
\bq
 \left( \bar{\phi}_\pm \right)^\dagger & = & \phi_\mp.
\eq
The asymptotic fields we may expand in terms of 
creation and annihilation operators:
For the scalars and the quarks we have:
\bq
\label{eq:asymptotic_expansion}
\phi_\pm(x) & = & \int \frac{d^3k}{(2\pi)^3 2 k^0} 
        \left( \hat{a}_\mp(k) e^{-i k x} + a_\pm^\dagger(k) e^{i k x} \right),
 \nonumber \\
 \Psi(x) & = & \sum\limits_{\lambda} \int \frac{d^3k}{(2\pi)^3 2 k^0}
    \left( b(k,\lambda) u(k,-\lambda) e^{-i k x} + d^\dagger(k,\lambda) v(k,\lambda) e^{i k x} \right).
\eq
Let us consider all particles in an amplitude as out-going. To derive supersymmetric relations among different
amplitudes with out-going particles we have to know 
through the reduction formulae 
the transformation laws of the annihilation operators, e.g
\bq
\lefteqn{
  A(\bar{Q}^\pm(k_1), Q^\pm(k_2), \dots, \bar{\phi}_\pm(k_j), \phi_\pm(k_{j+1}), \dots)
= } & & \nonumber \\
 & & 
 \vev{ b(k_1,\pm), d(k_2,\pm), \dots, \hat{a}_\pm(k_j), a_\pm(k_{j+1}), \dots}.
\eq
The annihilation operators can be projected out in eq. (\ref{eq:asymptotic_expansion}):
With the notation
\bq
 f \stackrel{\leftrightarrow}{{\partial}_\mu} g & = & f \left( \partial_\mu g \right) - \left( \partial_\mu f \right) g
\eq
we have
\bq
 a_\pm(k) = i \int d^3x \; \left( e^{i k x} \stackrel{\leftrightarrow}{{\partial}_0} \bar{\phi}_\mp(x) \right),
 & &
 \hat{a}_\pm(k) = i \int d^3x \; \left( e^{i k x} \stackrel{\leftrightarrow}{{\partial}_0} \phi_\mp(x) \right),
 \nonumber \\
b(k,\lambda) =  
 \int d^3x \; e^{i k x} \bar{u}(k,\lambda) \gamma^0 \Psi(x),
 & &
d(k, \lambda) = 
 \int d^3x \; e^{i k x} \bar{\Psi}(x) \gamma^0 v(k,\lambda).
\eq
These equations allow us to obtain the transformation laws of the annihilation operators from the known
transformation laws of the fields. For example
\bq
 \delta_\eta \hat{a}_-(k) 
 & = & 
 i \int d^3x \; \left( e^{i k x} \stackrel{\leftrightarrow}{{\partial}_0} \delta_\eta \phi_+(x) \right)
 = 
 i \int d^3x \; \left( e^{i k x} \stackrel{\leftrightarrow}{{\partial}_0} \delta_\eta \bar{\varphi}_-(x) \right)
 \nonumber \\
 & = &
 i \int d^3x \; \left( e^{i k x} \stackrel{\leftrightarrow}{{\partial}_0} \sqrt{2} \bar{\eta} \bar{\psi}_-(x) \right)
 = 
 i \int d^3x \; \left( e^{i k x} \stackrel{\leftrightarrow}{{\partial}_0} \sqrt{2} \bar{\eta} P_- \Psi(x) \right).
\eq
Reinserting the expansion eq. (\ref{eq:asymptotic_expansion}) for $\Psi(x)$ 
one finds the transformation laws
\bq
\delta_\eta a_\pm(k) & = & \sum\limits_\lambda d(k,\lambda) \left( \sqrt{2} \bar{v}(k,-\lambda) P_\pm 
  \left( \begin{array}{c} \eta \\ \bar{\eta} \\ \end{array} \right) \right),
 \nonumber \\
\delta_\eta \hat{a}_\pm(k) & = & \sum\limits_\lambda \left( \sqrt{2} \left( \eta, \bar{\eta} \right) P_\pm u(k,-\lambda) \right) b(k,\lambda).
\eq
For the fermion field we find
\bq
 \delta_\eta b(k,\lambda)
 & = & - \sqrt{2} \bar{u}(k,\lambda) \left( \begin{array}{c} \eta \hat{a}_-(k) \\ \bar{\eta} \hat{a}_+(k) \\ \end{array} \right),
 \nonumber \\
 \delta_\eta d(k,\lambda)
 & = & \sqrt{2} \left( a_-(k) \eta, a_+(k) \bar{\eta} \right) v(k,\lambda).
\eq
Using the explicit quark spinors in eq. (\ref{eq:os-spinors}) 
and taking the sign change for amplitudes with anti-quarks  discussed 
at the end of this section into account 
we obtain
eq. (\ref{eq:massive-susy}).
Similar considerations apply to the particles in the vector multiplet.
The gluino is a Majorana fermion with expansion
\bq
 \Lambda(x) & = & \sum\limits_{\lambda} \int \frac{d^3k}{(2\pi)^3 2 k^0}
    \left( b(k,\lambda) u(k,-\lambda) e^{-i k x} + b^\dagger(k,\lambda) v(k,\lambda) e^{i k x} \right).
\eq
The asymptotic expansion of the gluon reads
\bq
 A^\mu(x) & = & \sum\limits_{\lambda} \int \frac{d^3k}{(2\pi)^3 2 k^0}
   \left( a(k,\lambda) \eps^\mu(k,\lambda)^\ast e^{-i k x} + a^\dagger(k,\lambda) \eps^\mu(k,\lambda) e^{i k x} \right).
\eq
The annihilation operator for the gluon is projected out by
\bq
 a(k,\lambda) & = & 
 -i \int d^3x \; e^{i k x} \stackrel{\leftrightarrow}{{\partial}_0} \eps_\mu(k,\lambda) A^\mu(x).
\eq
We then obtain
\bq
\label{gluongluinotrafo}
  \delta_\eta a(k,\lambda) 
 & = & 
  \eps_\mu(k,\lambda)
 \left[ \left\l \eta - \left| \gamma^\mu \right| k - \right\r b(k,-) 
  - \left\l \eta + \left| \gamma^\mu \right| k + \right\r b(k,+)  
 \right],
 \nonumber \\
 \delta_\eta b(k,\pm) & = & 
 \pm \bar{u}(k,\lambda) \gamma^\mu P_\pm 
 \left( \begin{array}{c} \eta \\ \bar{\eta} \\ \end{array} \right)
 \sum\limits_{\lambda'} \eps_\mu^\ast(k,\lambda') a(k,\lambda').
\eq 
There is one subtlety in writing down the transformation laws for the fields:
For the application towards supersymmetric Ward identities we have to keep the relative
sign between different amplitudes correct.
Let us consider a Feynman graph corresponding to the matrix element
\bq
 \left\langle 0 \left| b ... d ...T\left[ \bar{\psi}_1 \Gamma_1 \psi_1 ... \bar{\psi}_n \Gamma_n \psi_n \right]
           b^\dagger ... d^\dagger \right| 0 \right\rangle.
\eq
Here $\bar{\psi}_j \Gamma_j \psi_j$ denotes a generic fermion interaction vertex.
In order to apply the Feynman rules, we have to anti-commute the fields and the creation and
annihilation operators such that
\bq
 \left\langle 0 \left| \psi \; b^\dagger(k,\lambda) \right| 0 \right\rangle \rightarrow u(k,\lambda), 
 & &
 \left\langle 0 \left| b(k,\lambda) \; \bar{\psi} \right| 0 \right\rangle \rightarrow \bar{u}(k,\lambda), 
 \nonumber \\
 \left\langle 0 \left| \bar{\psi} \; d^\dagger(k,\lambda) \right| 0 \right\rangle \rightarrow \bar{v}(k,\lambda), 
 & &
 \left\langle 0 \left| d(k,\lambda) \; \psi \right| 0 \right\rangle \rightarrow v(k,\lambda).
\eq
It is easily seen that this reordering brings a minus sign for each spinor $v$ and $\bar{v}$
\cite{Denner:1992vz}.
For an individual amplitude these signs are an overall factor and are usually ignored.
However, if one relates amplitudes with different particle contents 
through supersymmetric Ward identities to each other, these signs have to be taken into account.
We can incorporate these signs into the transformation laws, by adding an additional sign to each
transformation law, for which the number of anti-fermions changes by one unit.
For example
\bq
 \delta_\eta g^\pm 
  = - \Gamma^\pm_\eta(k) \bar{\Lambda}^\pm,
 & &
 \delta_\eta g^\pm 
  = \Gamma^\pm_\eta(k) \Lambda^\pm.
\eq
In the second equation we inserted an additional sign with respect to eq.(\ref{gluongluinotrafo}), which compensates for
the change in the number of anti-fermions.

%%%%%%%%%%%%%%%%%%%%%%%%%%%%%%%%%%%%%%%%%%%%%%%%%%%%%%%%%%%%%%%%%%%%%%%%%%%%
\section{Diagrammatic rules}
\label{sec:rules}

In~\cite{Schwinn:2005pi} we have
obtained diagrammatic rules involving only scalar propagators 
$i/k^2$ for gluons and $i/(k^2-m^2)$ for massive quarks. 
We introduce so called
 primitive vertices that are obtained by contracting
the vertices of the  standard colour ordered 
Feynman rules with the off-shell polarization vectors and spinors~\eqref{eq:os-pol} and~\eqref{eq:os-spinors}.
Within this approach it is convenient to take the momentum flow of a fermion always in the direction of the fermion
arrow line. That implies that we replace an outgoing anti-fermion by an incoming fermion.
All other particles remain outgoing.
The tri-valent primitive vertices  of massive quarks
 include vertices present both for massless
and massive quarks like
\bq
\label{eq:qqg}
   V_3(\bar{Q}_1^+,Q_2^-,g_3^+)= 
  i \sqrt 2 \frac{\sbraket{13}^2}{\sbraket{12}},
 & &
  V_3(\bar{Q}_1^-,Q_2^+,g_3^-)= 
  i \sqrt 2 \frac{\braket{31}^2}{\braket{21}}.
\eq
In addition, there are primitive
 vertices involving a helicity flip along the quark line
that vanish for massless quarks:
\bq
\label{eq:flip}
    V_3(\bar{Q}_1^+,Q_2^+,g_3^-)=
    i \sqrt 2 m \frac{\sbraket{12}^2}{\sbraket{23}\sbraket{31}},
 & &
     V_3(\bar{Q}_1^-,Q_2^-,g_3^+)=
    -i \sqrt 2 m \frac{\braket{12}^2}{\braket{23}\braket{31}}.
\eq
One can define 
the degree of a vertex or of an amplitude 
as the number of ``-''-labels minus one.
In the diagrammatic rules, only primitive vertices of degree zero and one 
occur.
Furthermore, the degree of an amplitude is exactly the sum of the degrees of the primitive vertices.

It is also straightforward to obtain the primitive vertices involving the
scalars and gluinos resulting from the Lagrangian~\eqref{eq:Lagrangian_2}.
For the scalar interactions with gluons we obtain the primitive tri-valent
vertices
\begin{equation}
\label{eq:susy_scalar}
\begin{aligned}
 V(\bar{\phi}_1^\pm,\phi_2^\mp,g_3^+)
& =i \sqrt 2\frac{\sbraket{13}\sbraket{23}}{\sbraket{12}},
&&
 V(\bar{\phi}_1^\pm,\phi_2^\mp,g_3^-)&=
  -i \sqrt 2\frac{\braket{13}\braket{23}}{\braket{12}}.
\end{aligned}
\end{equation}
The primitive vertices resulting from the couplings between quarks, gluinos
and scalars are obtained
by inserting the spinors of the massive quarks~\eqref{eq:os-spinors}.
One finds the nonvanishing tri-valent vertices
\bq
\label{eq:gluino-yukawa}
 V(\bar{Q}_1^+,\phi_2^-,\Lambda_3^+) = - i \sqrt{2} \sbraket{13},
 &&
 V(\bar{Q}_1^-,\phi_2^+,\Lambda_3^-) =  i \sqrt{2} \braket{13},
 \nonumber \\
 V(\bar{Q}_1^-,\phi_2^-,\Lambda_3^+) = - i \sqrt{2} m \frac{\braket{12}}{\braket{23}},
 &&
 V(\bar{Q}_1^+,\phi_2^+,\Lambda_3^-) =  i \sqrt{2} m \frac{\sbraket{12}}{\sbraket{23}},
 \nonumber \\
 V(\bar{\phi}_1^+,Q_2^-,\bar{\Lambda}_3^-) =  -i \sqrt{2} \braket{32},
 &&
 V(\bar{\phi}_1^-,Q_2^+,\bar{\Lambda}_3^+) =  i \sqrt{2} \sbraket{32},
 \nonumber \\
 V(\bar{\phi}_1^+,Q_2^+,\bar{\Lambda}_3^-) =  - i \sqrt{2} m \frac{\sbraket{12}}{\sbraket{13}},
 &&
 V(\bar{\phi}_1^-,Q_2^-,\bar{\Lambda}_3^+) =  i \sqrt{2} m \frac{\braket{12}}{\braket{13}}.
\eq
In addition there are four-valent vertices, which however we do not list here.
As for pure
QCD only degree one and zero vertices occur. For the application in
the SWIs it is important to note that the scalars
only couple to gluinos with the opposite helicity label.

%%%%%%%%%%%%%%%%%%%%%%%%%%%%%%%%%%%%%%%%%%%%%%%%%%%%%%%%%%%%%%%%%%%%%%%%%%%%
\section{Solution of the recursion relation}
\label{app:recursion}

In this appendix we give some details on the solution of the recurrence relations
(\ref{eq:Qrecurs}) and (\ref{eq:phirecurs}).
We follow closely the calculation of~\cite{Rodrigo:2005eu}.
For the solution to the recursion relation~\eqref{eq:Qrecurs}
we make the  ansatz
\begin{multline}\label{eq:ansatz}
  A_n(\bar Q_1^+,\dots,g_{n-1}^+|\widehat Q_{p_n}^-)=
\frac{A_n(\bar q_1^+,\dots,g_{n-1}^+|\widehat q_{p_n}^-)}{p_n^2 \left[(k_1+k_2)^2-m^2\right]}
\Bigl[(p_n^2-m^2)\braket{12}\sbraket{21}-m^2 B_n\Bigr],
\end{multline}
where  $A_n(\bar q_1^+,\dots,g_{n-1}^+|\widehat q_{p_n}^-)$ is the 
 amplitude
for \emph{massless} quarks with one
 leg off-shell~\cite{Berends:1987me,Mangano:1990by} 
\begin{equation}\label{eq:qcurrent++}
 A_n(\bar q_1^+,\dots,g_{n-1}^+|\widehat q_{p_n}^-)=
      2^{n/2-1} (-i k_{1,n}^2)
    \frac{\braket{p_nq}}{\braket{12}\dots\braket{(n-2) (n-1)}\braket{(n-1)q}}.
   \end{equation}
In the massless limit, the formula~\eqref{eq:ansatz}
 reduces to the known result~\eqref{eq:qcurrent++} while in
the on-shell limit a finite term proportional to $m^2$ remains that
is determined by the quantity $B_n$:
\begin{equation}
\label{eq:qc++_onshell}
 A_n(\bar Q_1^+,\dots,g_{n-1}^+|\widehat Q_{p_n}^-)=
\frac{i m^2 2^{n/2-1}}{(k_1+k_2)^2-m^2}
 \frac{\braket{p_nq}B_n}{\braket{12}\dots\braket{(n-1)q}}.
\end{equation}
 The structure of the
amplitude~\eqref{eq:ansatz} is similar to the current with two
off-shell gluons obtained in~\cite{Mahlon:1992fs}.

In the recursion relation~\eqref{eq:Qrecurs} also enters
the well known expression for the gluon
amplitude with one negative helicity gluon~\cite{Berends:1987me,Kosower:1989xy,Mangano:1990by,Dixon:1996wi} and one off-shell leg
\begin{equation}\label{eq:current++}
     A_n\left( g_1^+, ..., g_{n-1}^+,\widehat{g}_n^-\right)=
      2^{n/2-1} (-i k_{1,n}^2)
    \frac{\braket{ q n}^2}{\braket{q1}\braket{12}\dots\braket{(n-2) (n-1)}\braket{(n-1)q}}.
   \end{equation}
As a check, we have reproduced this amplitude from the Berends-Giele relations
 within the formalism of~\cite{Schwinn:2005pi}.

Inserting the ansatz~\eqref{eq:ansatz} into the recursion relation~\eqref{eq:Qrecurs},
 one obtains after some tedious steps a recursion relation for $B_n$ 
that is equivalent to that considered in
~\cite{Rodrigo:2005eu}, although there the part vanishing on-shell has been split off
differently.
 Adopting the solution given there, we
further split $B_n$ into two contributions
%sign changed on 09-01-2006 by CS
\begin{equation}
\label{eq:def_btilde}
  B_n=\frac{\braket{12}}{\braket{1q}(k_{1,4}^2-m^2)}\left[(k_n^2-m^2)
      \braket{2+|\fmslash k_3|q+}+\tilde B_n\right],
\end{equation}
with $\tilde B_4=0$. 
The function $\tilde B_n$ satisfies the recursion relation
\begin{equation}
  \label{eq:bt-recurs}
 \tilde B_n
 =\braket{2+|\fmslash k_1\fmslash k_{2,4}\fmslash k_{4,n}|q+}
+\sum_{j=5}^{n-1}\frac{\braket{j(j-1)}\braket{q-|\fmslash k_{1,j}\fmslash k_{j,n}|q+} }{\braket{(j-1)q}\braket{qj}}\frac{\tilde B_{j}}{(k_{1,j}^2-m^2)}.
\end{equation}
A closed solution to this equation can be found in~\cite{Rodrigo:2005eu}.
For $n=4$ we therefore obtain the result quoted in~\eqref{eq:ttgg}
while the on-shell amplitudes for $n\geq 5$ are given by
\begin{equation}
\label{eq:a++}
 A_n(\bar Q_1^+,\dots,g_{n-1}^+, Q_{n}^-)
=\frac{i m^2 2^{n/2-1}}{(k_{1,3}^2-m^2)(k_{1,4}^2-m^2)}
 \frac{\braket{nq}\tilde B_n}{\braket{1q}\braket{23}\dots\braket{(n-1)q}}.
\end{equation} 
The explicit results for the five and six-point functions obtained 
from~\eqref{eq:a++} and~\eqref{eq:bt-recurs} are given by
\begin{subequations}
\begin{align}
  A_5(\bar Q_1^+,g_2^+,g_3^+,g_4^+, Q_5^-)&=2^{3/3} i m^2
\frac{\braket{5q}}{\braket{1q}}
\frac{\braket{2+|\fmslash k_1\fmslash k_{2,4}|4-}}
{\braket{23}\braket{34}(k_{1,3}^2-m^2)(k_{1,4}^2-m^2)},\\
A_6(\bar Q_1^+,g_2^+,g_3^+,g_4^+,g_5^+, Q_6^-)&= 2^{2}i m^2 
 \frac{\braket{6q}}{\braket{1q}\braket{5q}\braket{23}\braket{34}\braket{45}
(k_{1,3}^2-m^2)(k_{1,4}^2-m^2)}\\
&\times \Bigl[
\braket{2+|\fmslash k_1\fmslash k_{2,4}\fmslash k_{4,6}|q+}
  -  \frac{\braket{q-|\fmslash k_{1,5}\fmslash k_5|4+}
\braket{2+|\fmslash k_1\fmslash k_{2,4}|4-}}{(k_{1,5}^2-m^2)}   \Bigr].
\nonumber
\end{align}
\end{subequations}
Comparing to the results for scalars obtained
in~\cite{Bern:1996ja,Badger:2005zh,Forde:2005ue} one finds that the
five point functions obviously satisfy the SWI~\eqref{eq:massive-swi},
while the agreement of the six point functions can be established after
some use of momentum conservation and Dirac algebra.

%%%%%%%%%%%%%%%%%%%%%%%%%%%%%%%%%%%%%%%%%%%%%%%%%%%%%%%%%%%%%%%%%%%%%%%%%%%%

%%%%%%%%%%%%%%%%%%%%%%%%%%%%%%%%%%%%%%%%%%%%%%%%%%%%%%%%%%%%%%%%%%%%%%%%%%%%%
\providecommand{\href}[2]{#2}\begingroup\raggedright\endgroup

\end{document}